\documentclass{article}
\usepackage{amssymb,latexsym}
\usepackage[dvipdf]{epsfig}
\usepackage{color}
\begin{document}
\def\be{\begin{equation}}
\def\ee#1{\label{#1}\end{equation}}

\title{Model for a Universe described by a non-minimally coupled scalar field and interacting dark matter}
\author{J. B. Binder\footnote{jbb01@fisica.ufpr.br}$\,$
 and G. M. Kremer\footnote{kremer@fisica.ufpr.br}\\
Departamento de F\'\i sica, Universidade Federal do Paran\'a\\ Caixa
Postal 19044, 81531-990 Curitiba, Brazil}
\date{}
\maketitle

\begin{abstract}
In this work it is  investigated the evolution of a Universe where a
scalar field, non-minimally coupled to space-time curvature, plays
the role of quintessence and drives the Universe to a present
accelerated expansion. A non-relativistic dark matter constituent
that interacts directly with dark energy is also considered, where
the dark matter particle mass is assumed to be proportional to the
value of the scalar field. Two models for dark matter pressure are
considered: the usual one, pressureless, and another that comes from
a thermodynamic theory and relates the pressure with the coupling
between the scalar field and the curvature scalar. Although the
model has a strong dependence on the initial conditions, it is shown
that the mixture consisted of dark components plus baryonic matter
and radiation can reproduce the expected red-shift behavior of the
deceleration parameter, density parameters and luminosity distance.

\end{abstract}

\section{Introduction}

The measurements of the rotation curves of spiral galaxies \cite{1}
as well as other astronomical experiments suggest that the luminous
matter represents only a small amount of the massive particles of
the Universe, and that the more significant amount is related to
dark matter.

Recently the astronomical observations with super-novae of type Ia
suggested that our Universe is presently submitted to an accelerated
expansion \cite{2,3}; the nature of the responsible entity, called
dark energy, still remains unknown. The simplest explanation for the
acceleration is a cosmological constant  (see \cite{4}), which fits
the present data very well but has some important unsolved problems.
Another  possibility is to introduce a scalar field $\phi \left(t
\right)$, which has been extensively studied by the scientific
community. By  considering a barotropic equation of state for the
scalar field $p_{\phi} = \omega \rho_{\phi}$, we have a constant
value of $\omega = -1$ for the cosmological constant model and a
variable $\omega \geq -1$ for a minimally coupled scalar field
model. However, the measured data from Hubble Space Telescope
\cite{6} states the restriction for $\omega =
-1.02^{+0.13}_{-0.19}$, whereas some recent observations \cite{7}
has drawn the attention to the viability of models where $\omega <
-1$, which would invalidate both cosmological constant and minimally
coupled scalar field descriptions. Some phantom field models were
proposed in order to contemplate this hypothesis \cite{8}, but they
face also some strong difficulties \cite{9}.

 In this work we consider a scalar field non-minimally coupled to
 space-time
curvature, which was investigated in \cite{10} and widely studied
recently, among others,  in the works \cite{11,12,13,14,15}. In a
more recent paper \cite{16}  the suitability of such models was
studied and it was shown that the Einstein field equations are
stable under some conditions for the constant that couples the
scalar field with the curvature scalar.

Aside several models for the dark sector interaction
\cite{17,18,19,dan} -- that could replace the cosmological constant
model in the case of future experimental tests -- we consider here a
direct coupling between dark matter and dark energy. In this model,
studied in \cite{20} and more recently in \cite{21}, the dark matter
particle mass is proportional to the value of the scalar field that
represents the dark energy. However, in this model it is necessary
to have some precautions, since by assuming the dark matter particle
mass to change with time, we could make the dark matter energy
density to become physically inconsistent at early stages. We
emphasize that in this work the full set of differential equations
are solved numerically, unlike the asymptotic approximation of the
work \cite{21}. The baryonic matter and radiation are described here
as usual non-interacting components, being the first
non-relativistic and pressureless. However, for the dark matter two
models are considered: the first one, by assuming the dark matter as
pressureless, and the second one, by using a thermodynamic theory in
order to relate the effects of the non-minimally coupling to the
dark matter pressure. All components will be described by a set of
field equations, and the resulting observables, -- i.e., the density
parameters, the decelerating parameter and the luminosity distance
-- which are obtained as solutions of the field equations will be
compared to the available data set in order to drawn the conclusions
about the viability of this model. We show that physically
acceptable solutions are obtained, if we chose initial conditions
restricting the rolling nature of the scalar field at early stages.
Furthermore, there exist some freedom parameters that will be
important to fit the data from incoming experiments. The work is
organized as follows: in section 2 the basic field equations for
each component, i.e. dark energy, dark matter, baryons and
radiation, are derived. The solutions for the pressureless dark
matter are found in section 3 for given initial conditions, coupling
and potential constants, whereas in section 4 we give the
corresponding solutions for the dark matter with non-vanishing
pressure. We close the work with section 5 where we address to some
remarks and summarize the results of previous sections. Units have
been chosen so that $8 \pi G = c = \hbar = k = 1$, whereas the
metric tensor has signature ($+,-,-,-$).

\section{Field equations}

We shall consider that the Universe is modeled as a mixture of a
scalar field which plays the role of dark energy, a dark matter
field and non-interacting baryons and radiation. Here we adopt a
more general scalar field which is non-minimally coupled to
curvature \cite{10}, and a dark matter field whose particle mass is
proportional to the value of the scalar field at each point
\cite{21}. The action for this model is written as:
 \be
 S = \int d^4 x \sqrt{-g} \left\{ {1 \over 2} (1 - \xi \phi^2) R + {1 \over 2} \partial_\mu \phi
  \partial^\mu \phi - V \left( \phi \right)
  + {1 \over 2} \partial_\mu \chi \partial^\mu \chi
 -{1 \over 2} \zeta^2 \phi^2 \chi^2 + \mathcal{L}_r + \mathcal{L}_b
 \right\},
 \ee{1}
 where $\phi$ and $\chi$ are the dark
energy and dark matter scalar fields, respectively, $V \left( \phi
\right)$ is the dark energy potential density, $\xi$ is the coupling
constant between dark energy and the curvature scalar $R$ and $\zeta
\phi$ is the dark matter particle mass, with $\zeta$ being a
constant. Furthermore, $\mathcal{L}_r$ and $\mathcal{L}_b$ denote
the lagrangian densities of the radiation and of the baryons,
respectively.

By taking the variation of the action (\ref{1}) with respect to
the metric tensor $g_{\mu\nu}$, it follows Einstein field
equations
 \be
 R_{\mu\nu}-{1\over2}Rg_{\mu\nu}=-{T_{\mu\nu}\over (1-\xi\phi^2)}.
 \ee{2}
 In the above equation the total energy-momentum tensor $T_{\mu \nu}$ of the sources of the
 gravitational field is a sum of the energy-momentum tensors of
 the baryons $T_{\mu \nu}^b$, radiation $T_{\mu \nu}^r$, dark matter $T_{\mu \nu}^{dm}$ and scalar
 field $T_{\mu \nu}^{\phi}$, i.e., $T_{\mu \nu} = T_{\mu \nu}^b + T_{\mu \nu}^r +
 T_{\mu \nu}^{dm} + T_{\mu \nu}^{\phi}$. The expressions for the
 energy-momentum tensors read
 \be
 T_{\mu \nu}^b = {2\over \sqrt{-g}} {\delta \sqrt{-g} \mathcal{L}_b
 \over \delta g^{\mu \nu}},\qquad
  T_{\mu \nu}^r = {2 \over \sqrt{-g}} {\delta \sqrt{-g}
 \mathcal{L}_r \over \delta g^{\mu \nu}},
 \ee{3}
  \be
 T_{\mu \nu}^{dm} = \partial_{\mu} \chi \partial_{\nu} \chi -
 \left( {1 \over 2} \partial_\sigma \chi \partial^\sigma \chi - {1
 \over 2} \zeta^2 \phi^2 \chi^2 \right) g_{\mu \nu},
 \ee{4}
 \be
 T_{\mu \nu}^{\phi} = \partial_{\mu} \phi \partial_{\nu} \phi -
 \left( {1 \over 2} \partial_\sigma \phi \partial^\sigma \phi - V
 \right) g_{\mu \nu}
  - \xi \left( \nabla_{\mu} \nabla_{\nu} \phi^2 -
 g_{\mu \nu} \nabla_{\sigma} \nabla^{\sigma} \phi^2 \right).
 \ee{5}

 We call attention to the fact that due to the inclusion of the coupling between the
 scalar field and the scalar curvature, the covariant divergence of the total energy-momentum
 tensor does not vanish. Indeed, one can obtain
 from (\ref{2}) by using Bianchi identities
 \be
 \nabla_{\nu}{T^{\mu\nu}}=2\xi\phi\left(R^{\mu\nu}-{1\over2}Rg^{\mu\nu}\right)\partial_\nu\phi,
 \ee{6}
  indicating that  the total energy-momentum tensor is not a conserved
  quantity. However, we shall assume that the covariant divergence of the energy-momentum
  tensors of the baryons and radiation vanish, i.e.,
  $\nabla_{\nu}{T_b^{\mu\nu}}=0$ and $\nabla_{\nu}{T_r^{\mu\nu}}=0$,
  since they are considered as non-interacting fields from the decoupling age up to the present time.

The evolution equations for the scalar and dark matter fields are
obtained from  Euler-Lagrange equations, yielding
 \be
 \nabla_{\sigma} \nabla^{\sigma} \phi + {d V \over d
 \phi} + \xi R \phi + \zeta^2 \phi\,\chi^2  = 0,\quad
 \nabla_{\sigma} \nabla^{\sigma} \chi + \zeta^2 \phi^2\chi  = 0.
 \ee{7}

 Henceforth, we shall restrict ourselves to the study of the consequences of the above
 model on a Universe which is isotropic, homogeneous  and spatially
flat, described by the Robertson-Walker metric:
 \be
 ds^2=dt^2-a(t)^2 \left( dx^2+dy^2+dz^2 \right),
 \ee{8}
 where $a\left( t \right)$ denotes the cosmic scale factor.

 For an isotropic and homogeneous Universe the most general representation for the total energy-momentum
 tensor in a comoving frame is given by $({T^\mu}_\nu)={\rm diag}(\rho,-p,-p,-p)$,
where the total energy density $\rho$ and pressure $p$ of the
mixture are given in terms of  the respective quantities for its
constituents, namely, $\rho = \rho_b + \rho_r + \rho_{dm} +
\rho_\phi$ and $p = p_b + p_r + p_{dm} + p_\phi$. From now on, we
shall assume that the baryons are non-relativistic particles so
that $p_b=0$ and that  the barotropic equation of state for the
radiation field $p_r=\rho_r/3$ holds.

Now for the Robertson-Walker metric, Einstein field equations
(\ref{2}) lead to the following  modified forms of the Friedmann
and acceleration equations
 \be
  H^2 = {\rho \over 3 \left( 1 - \xi \phi^2 \right)}, \qquad
{\ddot a \over a} = - {\rho + 3p \over 6 \left( 1 - \xi \phi^2
 \right)},
 \ee{9}
 respectively, due to the non-minimally coupling of the scalar field to the gravitational field.
 Above $H=\dot a(t)/a(t)$ is the Hubble parameter and the dot
denotes a differentiation with respect to time.

The hypothesis of homogeneity implies that the fields $\phi$ and
$\chi$ must be only functions of time so that the evolution
equations for the dark energy (\ref{7})$_1$ and dark matter
(\ref{7})$_2$ fields reduce to
 \be
 \ddot \phi + 3 H \dot \phi + {d V \over d \phi} + {\xi
 \phi \over 1 - \xi \phi^2} \left( \rho - 3p \right) + {1 \over \phi}
 \left( \rho_{dm} -p_{dm}  \right) = 0,
 \ee{10}
 \be
 \ddot \chi + 3 H \dot \chi +  \zeta^2 \phi^2\chi = 0,
 \ee{11}
respectively.

The energy densities and the pressures of the dark energy and  dark
matter fields can now be determined from (\ref{4}) and (\ref{5}),
yielding
 \be
 \rho_\phi = {1 \over 2}
 \dot \phi^2 + V + 6 \xi H \phi \dot \phi,
 \qquad
 p_{\phi}= {1 \over 2} \dot \phi^2 - V - 2 \xi \left( \phi \ddot
 \phi + \dot \phi^2 + 2H \phi \dot \phi \right),
 \ee{12}
 \be
 \rho_{dm} = {1 \over 2} \dot \chi^2 + {1 \over 2}\zeta^2 \chi^2
 \phi^2,\qquad
 p_{dm} = {1 \over 2} \dot \chi^2 - {1 \over 2}\zeta^2 \chi^2
 \phi^2.
 \ee{13}

 The evolution equation for the energy density of the scalar field
 is obtained by taking the time derivative of its energy
 density (\ref{12})$_1$  which, after some rearrangements, yields
 \be
 \dot \rho_\phi+3H(\rho_\phi+p_\phi)=-\left( \rho_{dm} - p_{dm} \right) { \dot \phi \over \phi}
 -{2\xi\phi\dot\phi \rho\over 1-\xi\phi^2},
 \ee{14}
 thanks to (\ref{10}) and (\ref{12}).
 Following the same methodology, the time derivative of the energy density
 of the dark matter (\ref{13})$_1$  leads  to the evolution equation for dark matter
 density:
 \be
 \dot \rho_{dm} + 3H \left( \rho_{dm} + p_{dm} \right) =
 \left( \rho_{dm} - p_{dm} \right) { \dot \phi \over \phi},
 \ee{15}
 by using equations (\ref{11}) and (\ref{13})$_2$.
From equations (\ref{14})  and (\ref{15}) we infer that there exists
a transfer of energy from the scalar field to the dark matter field
which is given by the term $(\rho_{dm}-p_{dm}) { \dot \phi / \phi}$.
Moreover, the term ${2\xi\phi\dot\phi \rho/( 1-\xi\phi^2)}$ in
(\ref{14}) is the responsible for the energy transfer from the
scalar field to the gravitational field.

One can obtain  from equation  (\ref{15}) together with the
definition $\rho_{dm} = m_{dm} n_{dm} = \zeta \phi n_{dm} $  a
general expression for the particle number density for dark matter
which reads
 \be
 \dot n_{dm} + 3 H n_{dm} = - {p_{dm} \over m_{dm}} \left(
 3H + {\dot \phi \over \phi} \right).
 \ee{15.2}
 The above equation clearly
simplifies to the usual $\dot n_{dm} + 3H n_{dm} = 0$ for the
vanishing dark matter pressure case and otherwise leads to a non
conserved dark matter particle number density due to the particle
creation by thermodynamic processes. In the next sections we shall
find the cosmological solutions for this model by considering two
cases, namely, a pressureless dark matter and a dark matter field
with non-vanishing pressure.

 The conservation of the energy-momentum tensor of baryons and radiation
with $p_b=0$ and $p_r=\rho_r/3$ lead to the well-known
relationships $\rho_b\propto 1/a^3$ and $\rho_r\propto 1/a^4$.

\section{Pressureless dark matter}

For the case of a pressureless dark matter we have $p_{dm}=0$ and it
follows from equation (\ref{15.2}) that $\rho_{dm}\propto \phi/a^3$.
Moreover, this last relationship implies  that the particle number
density of the dark matter is conserved, i.e., $\dot
n_{dm}+3Hn_{dm}=0$. Here we have only one equation to solve --
namely, the evolution equation for the scalar field (\ref{10}) -- in
order to obtain the time evolution of the acceleration field and of
the energy densities of each component. By changing the variables,
we can express the equation for the dark energy scalar field
(\ref{10}) in terms of the red-shift $z$, since $a =
1/\left(1+z\right)$  by assuming that  $a(0)=1$. Hence it follows
 \be
{ (1+z)^2 \rho \over 3 (1-\xi \phi^2 )}
 \phi''-{(1+z)(\rho-3p) \over 6 ( 1-\xi \phi
 ^2)}\phi'
 +{d V \over d \phi}
 +{\xi \phi ( \rho - 3 p )\over 1 -\xi \phi^2}
  + {\rho_{dm} \over \phi}= 0,
 \ee{16}
where hereafter the prime denotes the  differentiation with respect
to $z$.

In order to solve (\ref{16}) we have to choose a form for the
potential density. Among the several existent models (see, for
example~\cite{22}), we select a potential that was also used in the
references~\cite{20,21,24}:
 \be
 V \left( \phi \right) = {K \over
 \phi^{\alpha}},
 \ee{17}
where $K$ is a constant. This form  has the property of blowing up
for small values of $\phi$, preventing the field -- and therefore
the dark matter energy density --  from becoming negative.
Furthermore, we have to specify initial conditions for the scalar
field and its derivative as well as  for the energy densities of the
baryons, radiation and dark matter. The initial conditions for the
energy densities were chosen from the present known values given in
the literature (see \cite{25} for a review) for the density
parameters  $ \Omega_{i}(z) = {\rho_i(z) / \rho(z)}$, i.e.,
$\Omega_{b}(0) = 0.04995$, $\Omega_{r}(0) = 5\times10^{-5}$ and
$\Omega_{dm}(0) = 0.23$. The constant $K$ was determined from the
present value for the dark energy density parameter, i.e.,
$\Omega_{\phi}(0)= 0.72$, since it was considered that the present
value of the scalar field has a small value, so that
$\rho_{\phi}(0)\approx V$. There still remain $\alpha$ and $\xi$ as
free parameters, whose influence will be studied later in this
section.

\begin{figure}[htbp]
\begin{center}
   \includegraphics[width=7cm]{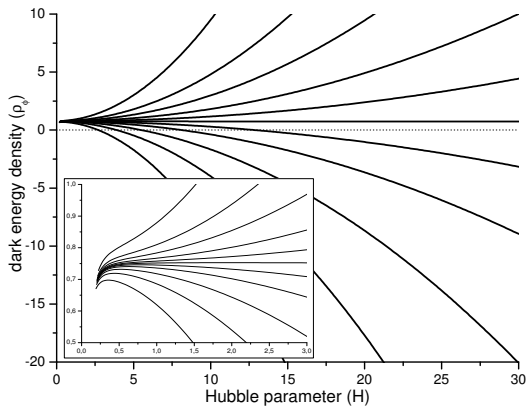}
   \caption[]{Typical phase-portrait ($\rho_{\phi},H$) corresponding to
the case $\alpha = 1/3$ and $\xi = -0.2$ showing the evolution for both
physical ($\rho_{\phi} > 0$) and non-physical ($\rho_{\phi} < 0$) solutions.\label{figure1}}
\end{center}
\end{figure}

We have plotted in figure \ref{figure1} a phase-portrait that exhibits the
evolution of a set of initial conditions. By varying the values of
the scalar field in order to produce small variations in the initial
dark energy density, it follows that some solutions evolve into a
non-physical behavior since  they lead to negative values of
$\rho_{\phi}$, whereas there exists another set of solutions that
causes the dark energy density to increase too fast and is the
responsible for almost all contributions to $H$ and $\rho$. In order
to get a physical acceptable behavior, it is necessary to impose a
very small and positive value for the scalar field slope at early
epochs, i.e.,  $z ^>_\sim 1000$. Physically this means that the
scalar field starts to roll more significantly around that time and
then evolves until today. Moreover, negative derivative of the
scalar field at high red-shifts would imply on negative dark energy
density, due to (\ref{12}).

\begin{figure}[htbp]
\begin{center}
   \includegraphics[width=6.2cm]{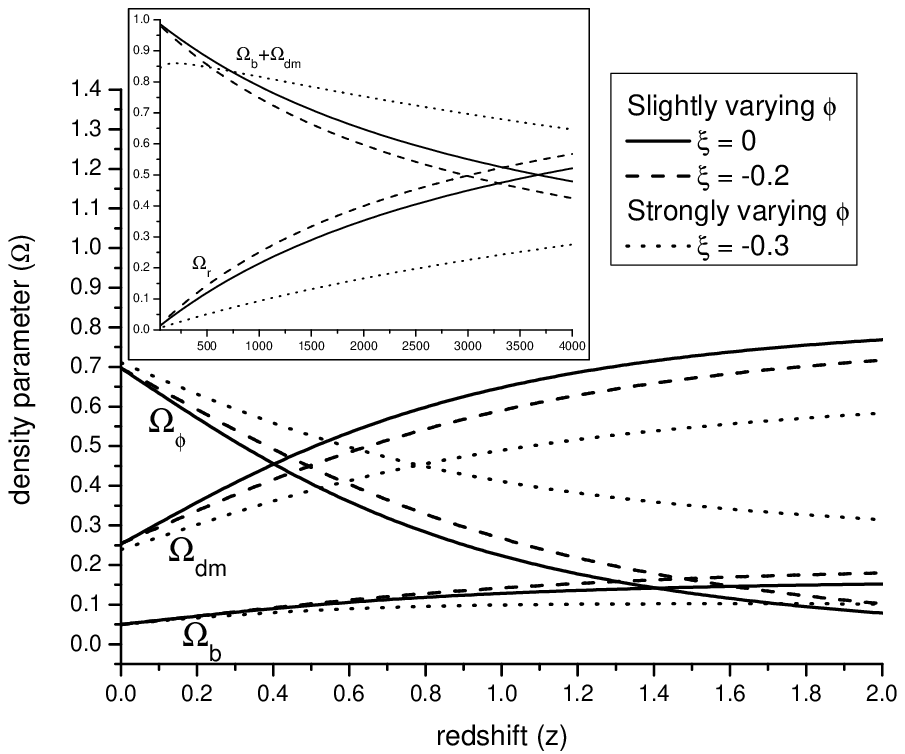}\hskip1.5cm
\includegraphics[width=6.2cm]{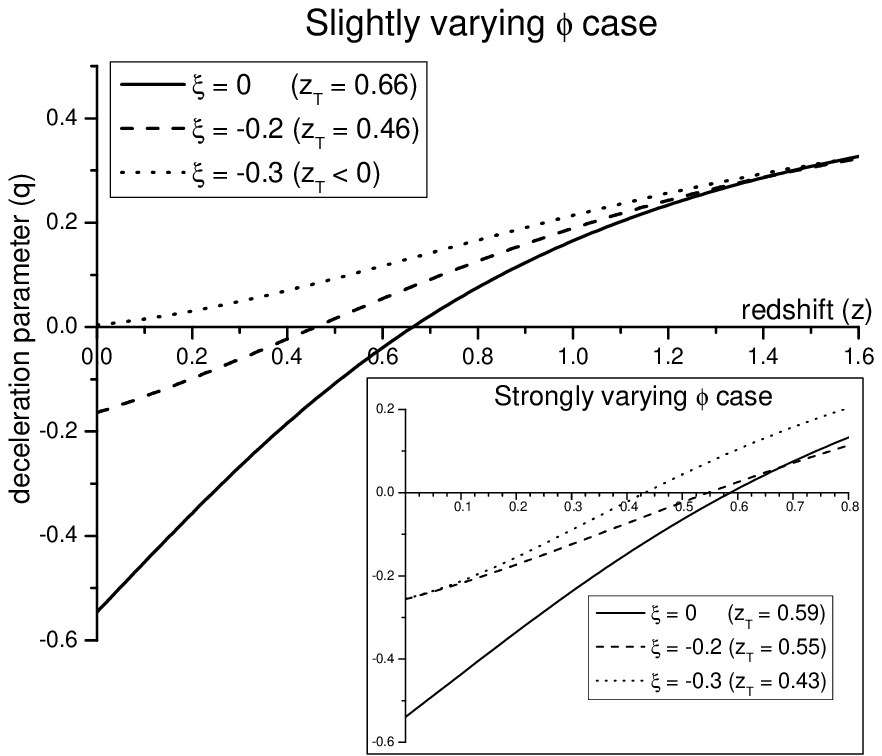}
\caption{Left: evolution of density parameters $\Omega_i$ vs.
red-shift $z$, showing that the coupling changes dark energy-dark
matter equality and radiation-matter equality with the red-shift;
right: deceleration parameter $q$ vs. red-shift $z$, the transition
red-shift is closer to $0$ as the coupling increases.
\label{figure2}}
\end{center}
\end{figure}

The red-shift evolution of the density parameters is plotted in the
left frame of figure \ref{figure2}. We have considered a fixed value for the
exponent $\alpha$, namely $\alpha = 1/3$, and different values for
the coupling constant $\xi$. Two different cases were analyzed: (a)
a slightly varying scalar field where the scalar field does not
change significantly with the red-shift, showing that the
differences from the usual quintessence model ($\xi = 0$) are
basically due to the influence of the non-minimal coupling ($\xi =
-0.2$); (b) a strongly varying scalar field,  characterized by
$\xi=-0.3$, where  the initial value of the scalar field has a
smaller value than that of the former case but it has a more
accentuated change when the red-shift increases.  We infer from this
figure that by increasing the red-shift the density parameter of the
dark energy decays more slowly for the strongly varying case
($\xi=-0.3$) followed by the slightly varying case when $\xi=-0.2$
and $\xi=0$, whereas the density parameter of the dark matter
increases more slowly for these cases. This can be understood by
noting that the last term of the right-hand side of equation
(\ref{14}) -- that depends on the coupling constant -- reduces the
energy transfer from the scalar to the dark matter field. For
$\xi=0$ the dark energy-dark matter equality occurs at
$z\approx0.4$, the dark energy-baryons equality at $z\approx 1.4$,
whereas for $\xi=-0.2$ and $\xi=-0.3$ these equalities occur at
higher red-shifts. Moreover, for $\xi=0$ the radiation-matter
(baryons plus dark matter) equality occurs at $z\approx3700$,
whereas for $\xi=-0.2$ and $\xi=-0.3$ this equality happens for
lower and higher red-shift values, respectively.

The deceleration parameter $q = - {\ddot a a / \dot a ^2}$ as a
function of the red-shift $z$ is plotted in the right frame of
figure \ref{figure2}. We note that by increasing the coupling constant the
transition from a decelerated to an accelerated phase occurs at
lower red-shifts. According to present experimental values~\cite{6}
the decelerated-accelerated transition  occurs at $z_T = 0.46 \pm
0.13$ and we can observe from figure \ref{figure2} that for the slightly varying
case with $\xi=-0.2$ this transition is at $z_T=0.46$, whereas for
the strongly varying case with $\xi=-0.3$ at $z_T=0.43$. However, in
both cases the present value of the deceleration parameter has a
smaller modulus than the one given in the literature for the
$\Lambda CDM$ case, i.e., $q_0\approx-0.55$.

By changing the values of the potential exponent between the range
$0\leq\alpha\leq5 $ it is not possible to see any significant
difference in the behavior of the curves in figure \ref{figure2}, however, for
larger values ($\alpha \approx 10$), there is a slower decay of the
dark energy density and a later transition to the decelerated phase.
This occurs because $V$ is of the same order as the other terms for
red-shifts up to $z \approx 4$, and in this range the potential is
mainly controlled by the value of $K$ for a slightly  varying
$\phi$.

\begin{figure}[htbp]
\begin{center}
   \includegraphics[width=7cm]{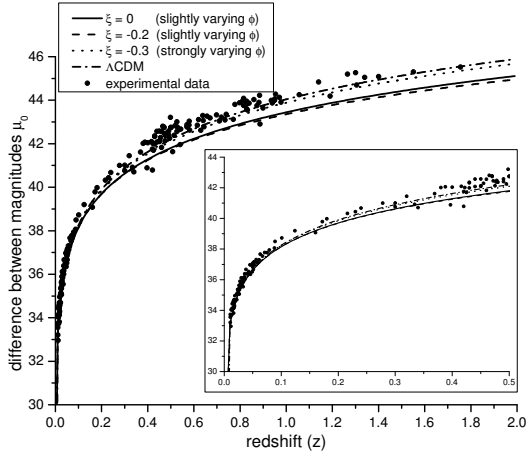}
   \caption[]{Difference $\mu_0$ vs. red-shift $z$, showing that coupling can increase or decrease
   the difference between magnitudes. \label{figure3}}
\end{center}
\end{figure}

In figure \ref{figure3} we have plotted the difference $\mu_0$ between the
 apparent magnitude $m$ and the absolute magnitude $M$ of a source,
given by $ \mu_0 = m - M = 5 \log d_L + 25$ with
 \be
 d_L = \left( 1+z\right) cH_0^{-1}
\int^z_0
 {dz \over H (z)},
 \ee{18}
  as a function of the red-shift $z$, where $d_L$ is the luminosity
distance given in Mpc.  In this figure the circles represent the
experimental values taken from the work by Riess \emph{et al.}
\cite{6} for 185 data points of super-novae of type Ia, whereas the
dash-dot line stands for the standard $\Lambda CDM$ model. We infer
from the figure that all cases fit well at low red-shifts, but only
the strongly varying case with  $\xi = -0.3$ stays near the
$\Lambda CDM$ over all the experimental range.

 To summarize, the best solutions are those
 that allow the scalar field to change only until the dark energy
density parameter becomes  small  and  the scalar field slope tends
to a small positive quantity.

\section{Dark Matter with non-vanishing pressure}

Let us analyze the case where the dark matter has a non-vanishing
pressure. For this end we recall that in the presence of matter
creation the first law of thermodynamics for adiabatic ($dQ=0$) open
systems reads~\cite{26}
 \be
 dQ=d(\rho V)+pdV-{\rho_{dm}+p_{dm}\over n_{dm}}d(n_{dm}V)=0,
 \ee{19}
where $V$ denotes the volume and it was supposed that only  dark
matter creation is  allowed. By considering $V\propto a^3$ and the
conservation laws for baryons and radiation it follows from
(\ref{19})
 \be
 \dot\rho_\phi+\dot\rho_{dm}+3H(\rho_\phi+\rho_{dm}+p_\phi+p_{dm})
 -{\rho_{dm}+p_{dm}\over n_{dm}}(\dot n_{dm}+3Hn_{dm})=0.
 \ee{20}
If we compare the above equation with the sum of equations
(\ref{14}) and (\ref{15}), we obtain
 \be
 {2\xi\phi\dot\phi\rho\over
 1-\xi\phi^2}=-{\rho_{dm}+p_{dm}\over n_{dm}}(\dot n_{dm}+3Hn_{dm}).
 \ee{21}
Now by using again (\ref{15}) and $\rho_{dm}=m_{dm}n_{dm}=\zeta\phi
n_{dm}$, we get the following  expression for the pressure of the
dark matter
 \be
 {p_{dm}\over\rho_{dm}}=-{1\over2}\pm\sqrt{{1\over4}+{2\xi\phi\dot\phi\rho\over
 (1-\xi\phi^2)(3H+\dot\phi/\phi)\rho_{dm}}}.
 \ee{22}
We call attention to the fact that only the plus sign must be chosen
in the above equation in order to obtain a positive dark matter
pressure. Moreover, we observe that in this case, when $\xi=0$ we
have from equation (\ref{21})  that the particle number density of
the dark matter is conserved, i.e., $\dot n_{dm}+3Hn_{dm}=0$ and
from (\ref{22}) that the dark matter pressure vanishes. For the case
analyzed in previous section there exists no constraint in the
values of $\xi$ when we assume from the beginning that the dark
matter pressure vanishes.

We have now a system of coupled differential equations to solve,
namely
 \be
 {(1+z)^2 \rho \over 3 (1-\xi \phi^2 )}
 \phi''-{(1+z)(\rho-3p) \over 6 ( 1-\xi \phi
 ^2)}\phi'+{d V \over d \phi}
  +{\xi \phi ( \rho - 3 p )\over 1 -\xi \phi^2}
  + { \rho_{dm} -p_{dm}\over \phi} = 0,
  \ee{23}
  \be
 \rho_{dm}^\prime - {3( \rho_{dm} + p_{dm}) \over (1+z)}=
 \left( \rho_{dm} - p_{dm} \right) { \phi^\prime \over \phi},
 \ee{24}
 where $p_{dm}$ is given by (\ref{22}) with the positive sign. The
 initial conditions used here are the same as those of the previous section.

\begin{figure}[htbp]
\begin{center}
   \includegraphics[width=6.2cm]{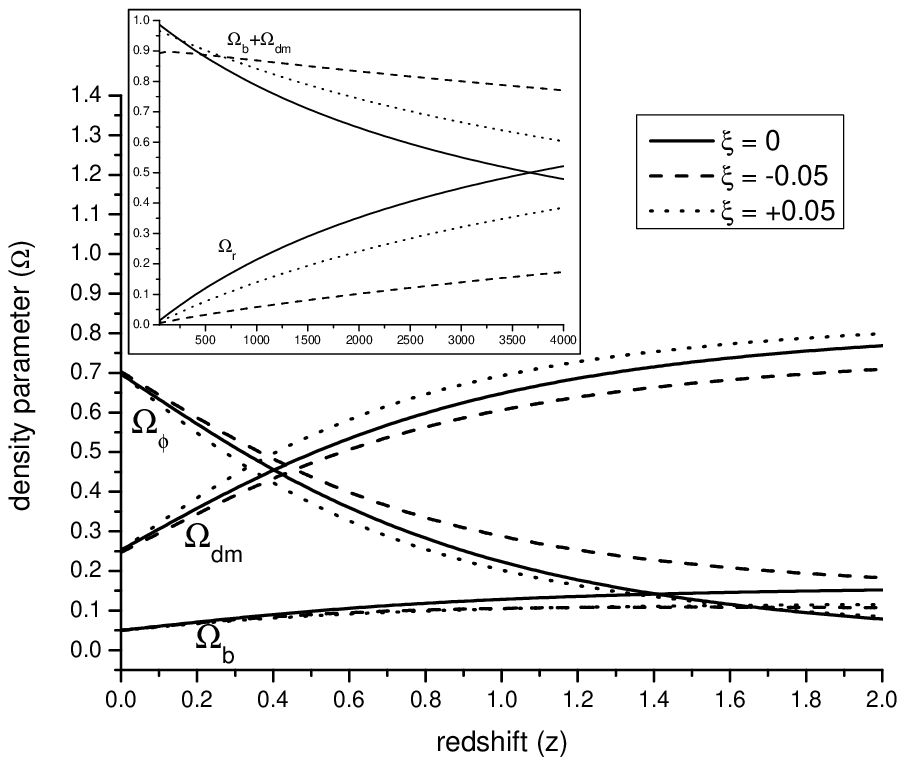}\hskip1.5cm
   \includegraphics[width=6.2cm]{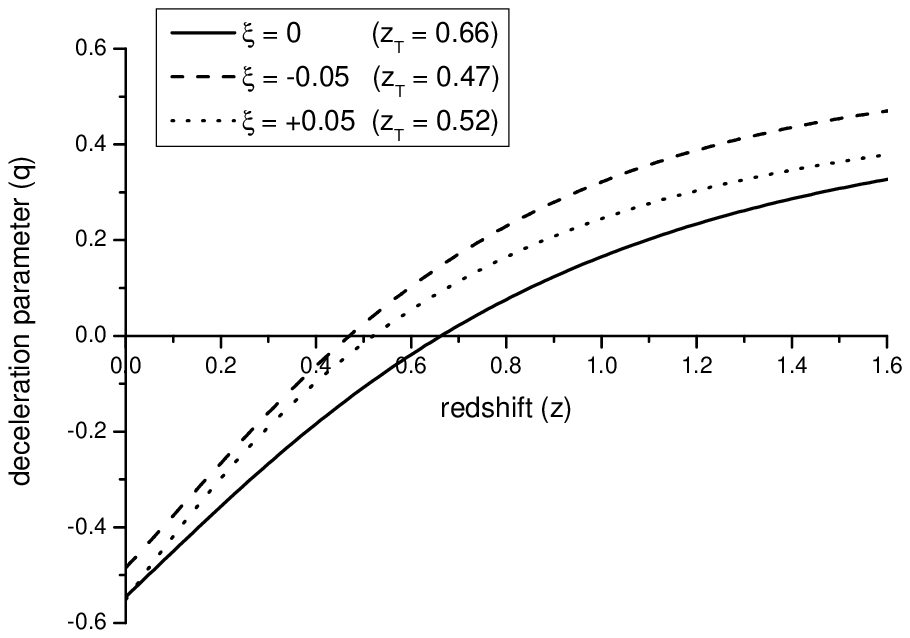}
\caption{Left: evolution of density parameters $\Omega_i$ vs.
red-shift $z$; right: deceleration parameter $q$ vs. red-shift $z$.
The uncoupled case represents an intermediate case between positive
and negative couplings for the density parameters at lower
red-shifts ($z<2$), and the coupling leads to a more recent
transition redshift.\label{figure4}}
\end{center}
\end{figure}

Figure \ref{figure4} corresponds to figure \ref{figure2} for the case of a non-vanishing
dark matter pressure, where we have analyzed two cases ($\xi =
\pm0.05$) and compared them with the minimally coupled pressureless
case. We observe from the left frame of this figure that  the decay
of the density parameter of the scalar field for the negative
coupling constant is slower than the one for the pressureless case,
whereas for the positive coupling constant the decay is more
accentuated. This can also be understood by analyzing the term
$-{2\xi\phi\dot\phi \rho/( 1-\xi\phi^2)}$ on the right-hand side of
evolution equation for the energy density of the scalar field
(\ref{14}), since it changes its sign when  $\xi$ is positive or
negative and it is equal to zero for the pressureless case. However,
for both cases the radiation-matter (baryons plus dark matter)
equality occurs for higher red-shifts than that of the pressureless
case. From the right frame of figure \ref{figure4} we infer that the transition
from a decelerated to an accelerated phase for $\xi=-0.05$ occurs at
$z_T=0.47$ and the present deceleration parameter for this case
approaches the value $q_0\approx-0.55$.

\begin{figure}[htbp]
\begin{center}
   \includegraphics[width=7cm]{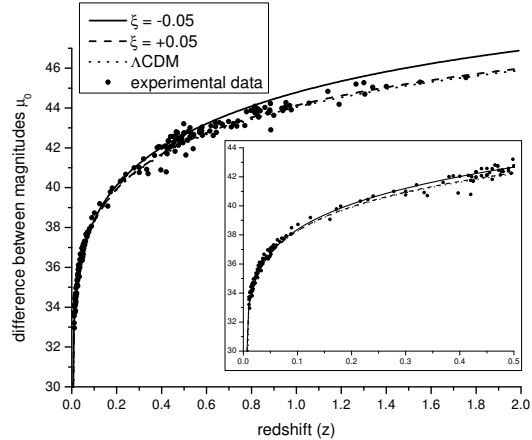}
\caption[]{Difference $\mu_0$ vs. red-shift $z$ where the  coupling
may increase the difference between magnitudes.\label{figure5}}
\end{center}
\end{figure}

In figure \ref{figure5} it is plotted the difference between the
apparent magnitude and the absolute magnitude  of a source
as a function of the  red-shift for the case of a non-vanishing dark
matter pressure. We observe that the $\xi = +0.05$ case stays near
the $\Lambda CDM$ model, whereas the $\xi = -0.05$ case improves the
behavior of the theoretical curve with respect to the experimental
values for low red-shifts.

\begin{figure}[htbp]
\begin{center}
   \includegraphics[width=6.2cm]{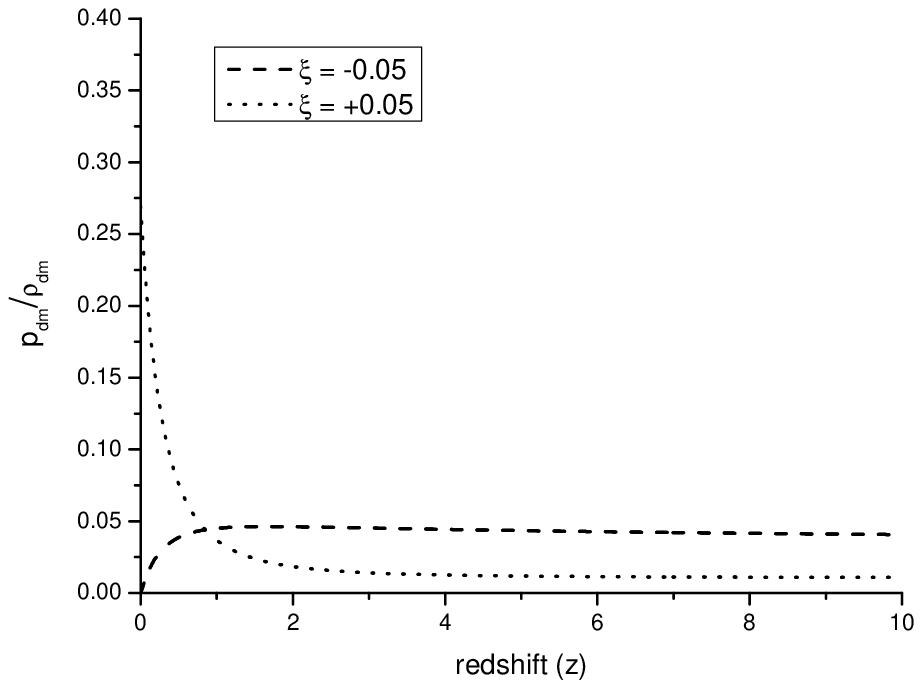}\hskip1.5cm
   \includegraphics[width=6.2cm]{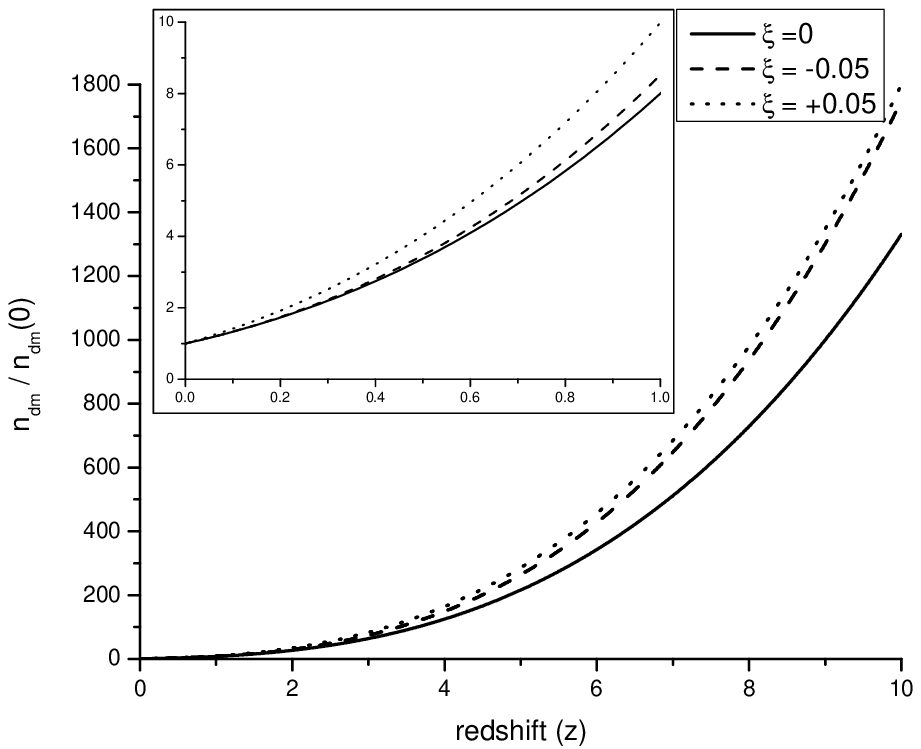}
\caption[]{Left: dark matter pressure $p_{dm}/\rho_{dm}$ vs.
red-shift $z$; right: dark matter particle number density
$n_{dm}(z)/n_{dm}(0)$ vs. red-shift $z$, showing the influence of
the coupling into the pressure and particle number density of the
dark matter.\label{figure6}}
\end{center}
\end{figure}

Finally, the left frame of figure \ref{figure6} shows the evolution of the dark
matter pressure for the situations discussed above. We note that for
$\xi=-0.05$ the dark matter pressure vanishes when the red-shift
goes to zero and tends to a small value for higher red-shifts. We
emphasize that although low, the dark matter pressure introduced
here has some significant influence on the behavior of the evolution
of the whole system. For completeness we have plotted in the right
frame of this figure the evolution of the dark matter particle
number density as a function of the red-shift, and, as was expected,
the particle number densities for the non-vanishing pressure cases
increase more rapidly with the red-shift than that of the
pressureless case.

\section{Final remarks and conclusions}

 An equivalent description for the model studied in previous
 sections is to write Einstein field equations as
 \be
 R_{\mu\nu}-{1\over2}Rg_{\mu\nu}=-{{\cal T}_{\mu\nu}}.
 \ee{25}
In this case, the energy-momentum tensor of the sources is
conserved, i.e., $\nabla_\nu {\cal T}^{\mu\nu}=0,$ whereas the
energy-momentum tensor of the scalar field reads
 \be
 {\cal T}_{\mu \nu}^{\phi} = \partial_{\mu} \phi \partial_{\nu} \phi -
 \left( {1 \over 2} \partial_\sigma \phi \partial^\sigma \phi - V
 \right) g_{\mu \nu} - \xi \left( \nabla_{\mu} \nabla_{\nu}
 \phi^2-
 g_{\mu \nu} \nabla_{\sigma} \nabla^{\sigma} \phi^2 \right)
 +\xi\phi^2\left( R_{\mu\nu}-{1\over2}Rg_{\mu\nu}\right).
 \ee{26}

 By considering the Robertson-Walker metric and the representation for
 the energy-momentum tensor  $({{\cal T}^\mu}_\nu)={\rm
diag}(\tilde\rho,-\tilde p,-\tilde p,-\tilde p)$ it follows the
Friedmann and acceleration equations in their usual forms, i.e.,
 \be
   H^2 = {\tilde\rho \over 3 }, \qquad
{\ddot a \over a} = - {\tilde\rho + 3\tilde p \over 6 }.
 \ee{27}
However, the energy density and the pressure of the scalar field in
this description have
 new contributions that are given by the underlined terms:
\be
 \tilde\rho_\phi = {1 \over 2}
 \dot \phi^2 + V + 6 \xi H \phi \dot
 \phi+\underline{\xi\phi^2\tilde\rho}\,,
 \ee{28a}
 \be
 \tilde p_{\phi}= {1 \over 2} \dot \phi^2 - V - 2 \xi \left( \phi \ddot
 \phi + \dot \phi^2 + 2H \phi \dot \phi \right)+\underline{\xi\phi^2\tilde
 p}\,,
 \ee{28}
so that the sums
 $\tilde \rho=\tilde \rho_\phi+\rho_{dm}+\rho_b+\rho_r$ and
 $\tilde p=\tilde p_\phi+p_{dm}+p_b+p_r$ give
 \be
 \tilde\rho={\rho_\phi+\rho_{dm}+\rho_b+\rho_r\over
 1-\xi\phi^2}={\rho\over 1-\xi\phi^2},
\qquad
 \tilde p={p_\phi+p_{dm}+p_b+p_r\over 1-\xi\phi^2}={p\over
 1-\xi\phi^2},
 \ee{29}
showing the equivalence between this description and the former one.

To summarize, we first observe  that this model  has a strong
dependence on the initial conditions. For the pressureless dark
matter case we have shown that: (a) a stronger coupling between the
scalar field and the scalar curvature leads to a
decelerated-accelerated transition and matter-radiation equality at
lower red-shifts;   (b) a stronger scalar field variation implies
that  stronger couplings adjust better with the experimental data of
luminosity distance, but the matter-radiation equality occurs at
higher red-shifts. For the case of a non-vanishing dark matter
pressure we have shown that negative small values of the coupling
constant leads to: (a) a decelerated-accelerated transition at lower
red-shifts with a better adjustment of the present value of the
deceleration parameter, but the matter-radiation equality occurs at
higher reds-shifts; (b) a better adjustment with the experimental
data of luminosity distance at low red-shifts.

\section*{Acknowledgment}
We are indebted to  Prof. E. Gunzig  for calling our attention to
the case analyzed in the final comments.




\end{document}